# Diffusion-free photon upconversion driven by intramolecular triplet-triplet annihilation in engineered conjugated chromophores.

*Sara Mattiello[1], Sara Mecca[1], Alessandra Ronchi[1], Adiel Calascibetta[1], Giuseppe Mattioli[2], Francesca Pallini[1], Francesco Meinardi[1]\*, Luca Beverina[1]\* and Angelo Monguzzi[1]\**

[1] Dipartimento di Scienza dei Materiali, Università degli Studi Milano-Bicocca, via R. Cozzi 55, 20125 Milano – ITALY

[2] CNR - ISTITUTO DI STRUTTURA DELLA MATERIA Via Salaria Km 29,300 - C.P. 10 I-00015 - Monterotondo Scalo (RM), ITALY

E-mail: angelo.monguzzi@unimib.it

**Abstract**

The photon upconversion based on sensitized triplet-triplet annihilation (*s*TTA-UC) is a spin-flip mechanism exploited to recover the energy of dark triplet states in conjugated systems. In this process a high-energy fluorescent singlet is created through the collision and fusion of two low-energy triplets belonging to different diffusing molecules. Its excellent yield in solution under low excitation intensity and non-coherent light highlighted the huge potential of *s*TTA-UC to provide a breakthrough in solar technologies. However, its diffusion-limited bi-molecular nature limits its efficiency in the solid state. To overcome this issue, we designed a molecular systems able to host simultaneously more than one triplet, thus enabling a diffusion-free intramolecular TTA. We obtain the first direct demonstration of intramolecular triplet fusion by tailored photoluminescence spectroscopy experiments, thus opening the way to realize a new family of single molecule upconverters with huge potential in solar and lightening technologies by accessing the natural triplets energy reservoir.

**Introduction.**

Organic conjugated materials feature unique interactions with near-ultraviolet, visible and near-infrared photons, thus enabling discovery of new fundamental physics as well as sustainable technological applications. Very recent literature demonstrates their huge potential to manage the energy of the absorbed photons by exploiting spin-flip mechanisms involving triplet states like singlet fission, thermally activated delayed fluorescence, and intersystem crossing (ISC) enhanced phosphorescence. Indeed, the most recent advancement in both molecular materials and organic devices are related to triplet mediated processes capable of boosting the efficiency of silicon solar cells,[1] enabling the fabrication of vivid and low-consuming OLEDs [2-4] as well as cheap, light and easy-to-handle screens for broad use in daily non-invasive diagnostics[5,6]. Innovative photochemistry was also reported.[7]

Among spin-flip photophysical processes, the mechanism of photon upconversion based on sensitized triplet-triplet annihilation ($s$TTA-UC) has been extensively investigated in multicomponent solution systems in the last decade. As shown in Fig.1A, generally in this case the photon upconversion is the result of the fusion of the metastable long living triplet state of two different annihilator/emitter molecules upon diffusion-mediated collision, i.e. an intermolecular TTA, which results in the formation of a high-energy singlet excited state that decays radiatively. The emitter triplets are populated via energy transfer (ET) from the triplets of a low-energy absorbing molecule, that is, a light-harvester/sensitizer. [8-10] Due to its excellent yield in optimized systems surpassing 30% - close to the thermodynamic limit of 50% - under low excitation intensity comparable to the solar irradiance, [11] and considering its efficiency also under non-coherent light, the $s$TTA-UC surpasses the limitations of traditional photon upconversion mechanisms, such as two-photon absorption or sequential excited state absorption, that requires coherent and/or high-intensity radiation to be efficient. [12,13] For such reasons, the $s$TTA-UC is intensively investigated to provide a breakthrough in solar technologies, [14-16] low power bio-imaging, [17,18] optogenetics, [19] anticounterfeiting, [20] and oxygen sensing applications. [21] Nevertheless, achieving $s$TTA-UC in the solid state - which is more technologically manageable to make devices - is still an open challenge. Indeed, the TTA mechanism is a diffusion limited bi-molecular process, thus its effectiveness in solids is hindered by several factors such as poor excitons mobility, aggregation and phase segregation effects, and excited state quenching. Many efforts have been concentrated in developing solid upconverters, with encouraging results obtained with nanoparticles, [22,23] nanocrystals, [24,25] polymeric films, [26,27] co-crystals [28] and multilayer devices, all of them able to partially overcome the triplets diffusion limits. [29-31] In particular, the confinement in colloidal nanostructures of both the sensitizer and emitter components, is a winning approach in mitigating the diffusion dependency of the TTA yield. [32] Pushing to the limits this approach, several groups are now investigating a potential intramolecular TTA mechanism. In this case, two triplet excitons are simultaneously created on the same annihilator, thus the TTA can occur without exciton diffusion (Fig.1B). Several examples of supramolecular annihilators have been recently proposed to achieve intramolecular TTA. [33-36] Even though a clear cut demonstration is still missing, the reported evidences are building up strong support to its existence.

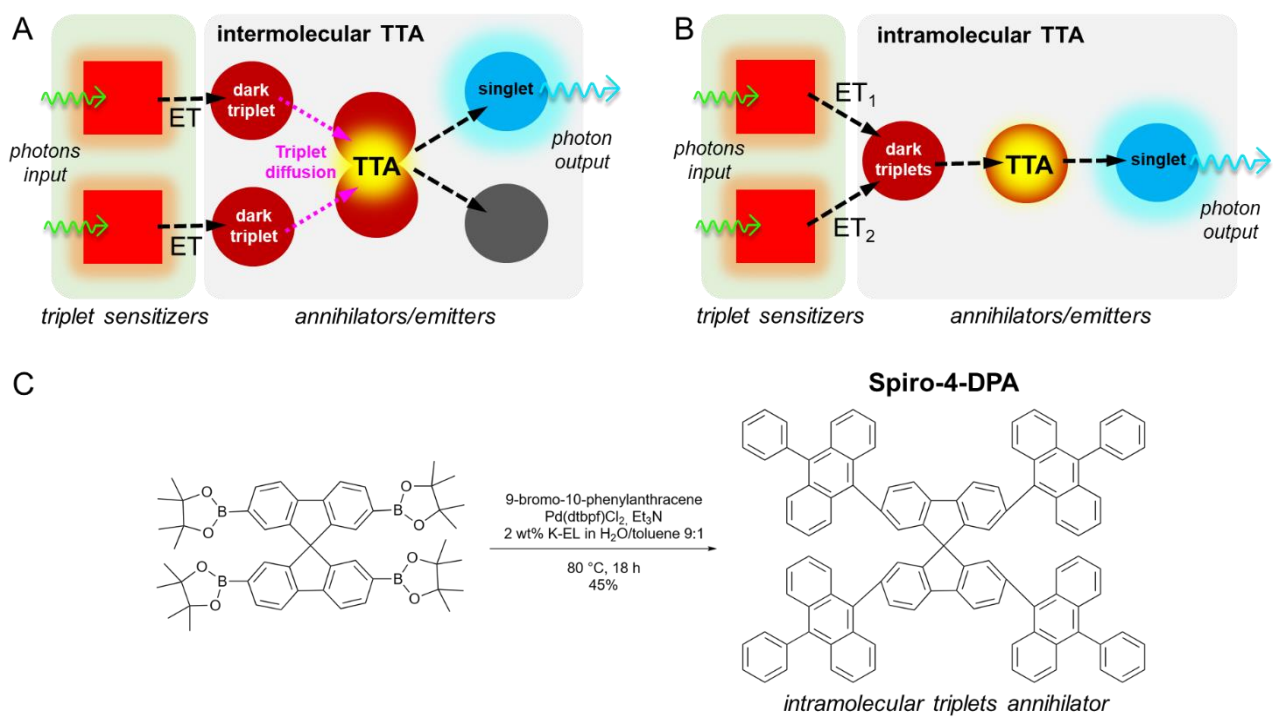

**Fig. 1 | Principles of photon upconversion based on sensitized triplet-triplet annihilation (sTTA-UC).** (**A**) In the standard intermolecular upconversion mechanism, the dark triplets of two emitter moieties are excited independently by energy transfer (ET) from excited triplet sensitizers that absorb low energy photons (green arrows). During their random diffusion, the two emitters can collide end experience intermolecular TTA generating an emitting singlet state from which upconverted emission is produced (blue arrow). (**B**) If two triplets can be generated on the same emitter by a subsequent two-step energy transfer (ET$_1$ and ET$_2$), the intramolecular TTA process can occur without the need of any diffusion of the annihilating triplets. (**C**) Synthetic route and molecular structure of the conjugated Spiro-4-DPA system designed for intramolecular TTA.

We report here for the first time the direct observation of intramolecular triplet fusion by TTA (*intra-TTA*) in a single molecular conjugated acceptor *2,2',7,7'-tetrakis(10-phenyl-anthracen-9-yl)-9,9'-spirobifluorene* (Spiro-4-DPA). The system has been designed to allow the simultaneous co-existence of more than one triplet, thus enabling the intramolecular TTA process. As shown in Fig.1B, upon a first energy transfer event ($ET_1$) from a sensitizer, one of the equivalent triplets is populated while the remaining triplet orbitals can host a second energy transfer event ($ET_2$) thus creating simultaneously two triplet excitons on the same molecule and enabling the intramolecular TTA without the need of the annihilating triplet diffusion. The demonstration of this intramolecular mechanism is obtained by steady state and time resolved photoluminescence spectroscopy experiments on the upconverted emission, finely tailored to clearly distinguish intra- from intermolecular TTA because of the inherently different dynamics of the two phenomena. Our results unequivocally demonstrate for the very first time the possibility of achieving multiple triplet excitations in the same molecule within a time span compatible with single molecule triplet spontaneous lifetime. This is a crucial step towards the realization of single molecule upconverters that could have disrupting applications in solar and lightening technologies profiting from the energy naturally stored in the omnipresent metastable triplets of conjugated systems.

**Results.**

The Spiro-4-DPA system has been synthesized by a micellar catalyzed Suzuki-Miyaura coupling between 2,2′,7,7′-tetrakis(4,4,5,5-tetramethyl-1,3,2-dioxaborolan-2-yl)-9,9′-spirobifluorene and 9-bromo-10-phenylanthracene performed in water solution of the industrial surfactant Kolliphor EL (K-EL) in the presence of a small amount of toluene. The product can be isolated by filtration and further purified by crystallization from toluene. In the overall the process is simple, efficient and requiring minimal amounts of organic solvents. Fig. 1C summarize the synthetic route exploited and the molecular structure of the conjugated annihilator Spiro-4-DPA. Synthesis details and structural characterization are reported in the Supplementary Material file (Supplementary Figures S1-S5). In optimized geometry, the use of the *spiro* linking center that exploits a compact $sp^3$ hybridized carbon atom makes the center-to-center distance between DPA lobes as short as 1.1 nm, thus fully enabling a localized TTA process.

Figure 2A shows the calculated transition densities to unravel the spatial distribution of excited states. [37][38] In particular, we focus on changes on the $T_1$ tripled state of Spiro-4-DPA by comparing the transition density just after the first ET (top), when the unrelaxed molecule retains the symmetrical structure of its ground state, to that calculated after the structural relaxation of the same excited state (bottom). Accordingly, panel B reports the calculated energy level diagram for the upconversion system investigated, where the Pt(II)-octaethyl porphyrin (PtOEP) is employed as triplet sensitizer.[23] Transition densities and electronic energies are calculated in the framework of time-dependent density functional theory (TD-DFT), as detailed in the Supplementary Material file.[39,40] Computational results are in excellent agreement with experiments, yielding a ground state absorption maximum at 3.16 eV (393 nm) with respect to the measured 3.11 eV (398 nm), as well as a fluorescence peak predicted at 2.95 eV (421 nm) vs. the 2.92 eV (424 nm) observed (Fig.S6). Control calculations performed on the reference 9,10-diphenylanthracene (DPA) molecule further confirm the reliability of the quantum mechanical modelling employed also for the estimation of the triplet state energies (Figs.S6, S7). The obtained results point out a crucial feature of the system, a particularly weak coupling between the four equivalent DPA units in the ground state. As a result, the triplet state $T_1$ is delocalized over the whole molecule (Fig.2A, top) but its energy of 1.75 eV its only marginally different from that of 1.77 eV of an isolated DPA (Supplementary Table S1). The excitation energy transfer ($ET_1$) from the PtOEP triplet at 1.94 eV to $T_1$ is thus energetically favorable in both cases. Now, it is worth noting that the Spiro-4-DPA system belongs to the S4 symmetry group (Supplementary Table S2), which does not imply any intrinsic fourfold degeneracy of orbitals.[41] However, in the symmetrical ground state the frontier orbitals display a close accidental fourfold degeneracy, which is reflected on excited singlet ($S_1$-$S_4$) and triplet ($T_1$-$T_4$) states, because of the presence of four weakly coupled identical lobes. The key feature of the excited states manifold of the Spiro-4-DPA with respect to DPA is that, after the first excitation transfer ($ET_1$), the populated $T_1$ localizes in one of the four lobes and its energy relaxes from 1.75 eV to 1.38 eV (Fig,2A, bottom). This process also lifts the degeneracy of the $T_2$-$T_4$ states, yet their energies remain in a limited 1.74 -1.76 eV range (Fig.2B). Such a rearrangement makes energetically feasible the transfer of a second excitation ($ET_2$) from sensitizers to any one of the $T_2$-$T_4$ states during the $T_1$ lifetime, while preserving the energetic constrain $T_1+T_{2(3,4)} > S_1$ pivotal to enable the formation of a singlet upon TTA. Moreover, given that the Spiro-4-DPA van der Waals diameter is 2.82 nm (Fig.S6), two neighboring DPA lobes simultaneously hosting triplet excitons are within the ~1 nm distance typically required for the short range TTA mechanism. [42] Therefore, the proposed system presents all the requirements to observe spontaneous intramolecular TTA process without the need of triplet excitons diffusion, thus overcoming the process limitations in the solid state discussed above.

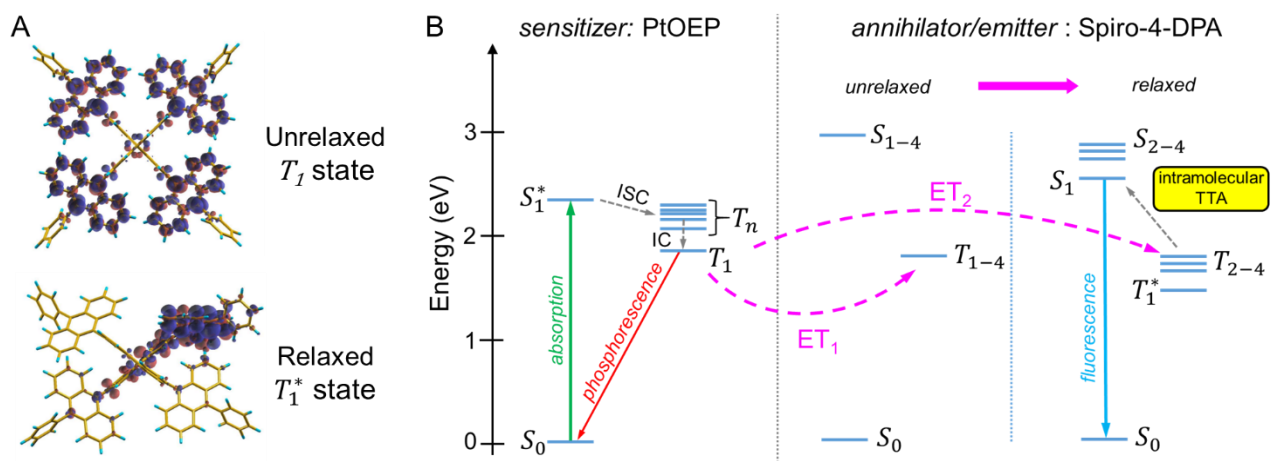

**Fig. 2 | Photophysics of photon upconversion based on intramolecular triplet-triplet annihilation (TTA).** (A) Transition densities for the unrelaxed ($T_1$, ground state molecule) and relaxed ($T_1^*$, excited molecule) triplet state of the conjugated Spiro-4-DPA molecule proposed as annihilator/emitter. (B) Energy-level diagram and sketch of the energy flux for the intramolecular TTA upconversion process. Dashed arrows mark radiationless transitions. The sensitizer molecule PtOEP is excited into a singlet state $S_1^*$ that undergoes intersystem crossing (ISC) into its triplet state manifold $T_2$-$T_6$. The system then relaxes by non-radiative internal conversion (IC) to the lower triplet state $T_1$. Energy transfer ($ET_1$) then occurs to the annihilator/emitter triplet $T_1$, which is 4-fold degenerate in the ground state. For clarity, the competitive back-ET to sensitizers triplets is omitted here due to its negligible efficiency. The emitter molecule then relaxes to a distorted excited state configuration where three triplet states ($T_2$, $T_3$, $T_4$) are still available to harvest a second excitation from sensitizers by $ET_2$. If the two triplets are simultaneously present on a single emitter molecule, they fuse by TTA and generate a high-energy singlet state $S_1$, whose spontaneous recombination to the ground state produces the upconverted fluorescence.

Preliminarily to the upconversion experiments, we investigated the electronic and emission properties of Spiro-4-DPA in tetrahydrofuran (THF) diluted solution ($10^{-5}$ M) by means of steady-state and time-resolved photoluminescence spectroscopy. As reported in Fig. 3A, with respect to the reference DPA standard, the Spiro-4-DPA shows a slightly red shifted absorption spectrum in the blue-near UV spectral region peaked at 389 nm, accompanied with a clearly resolved vibronic replica series at 378, 360 and 342 nm. Under UV excitation the system shows a bright blue photoluminescence, peaked at 424 nm, with an excellent quantum efficiency $\phi_{pl}$ of 0.79±0.08 and a characteristic emission lifetime of 3.2 nanoseconds (Fig.S8). These findings confirm the absence of strong coupling between the anthracene-like lobes of the Spiro-4-DPA, thus avoiding significant modification of their conjugation length and allowing to retain the excellent emission properties of the parent molecule DPA ($\phi_{pl}$ = 0.96). We tested the Spiro-4-DPA performances as annihilator/emitter for sTTA-UC in a $10^{-2}$ M THF solution with PtOEP ($10^{-4}$ M) as triplet sensitizer (Fig.S10). Upon continuous wave excitation with a 532 nm laser, the bicomponent solution shows a bright blue upconverted photoluminescence (Fig.3B, inset of panel 3D, Fig. S11). A minimal residual phosphorescence from PtOEP can be still observed at 645 nm, in agreement with the measured energy transfer yield $\phi_{ET}$ of 95% (Fig.S12). The occurrence of upconversion by intermolecular TTA is demonstrated by the time resolved photoluminescence experiments shown in Fig.3C. Under modulated excitation at 532 nm, the photoluminescence intensity at 430 nm shows a slow decay in the hundreds of microseconds time scale, with a characteristic lifetime that increases by reducing the excitation intensity. This behavior mirrors the generation of fluorescent singlets by intermolecular TTA.[43] Specifically, at the lowest excitation intensity employed of 0.1 mW cm$^{-2}$, the emission intensity decays as a single exponential function with characteristic lifetime $\tau_{uc}$. This means that in this excitation regime the TTA is a negligible deactivation channel for the Spiro-4-DPA triplets, thus the spontaneous decay time of the $T_1$ state can be directly calculated as twice $\tau_{uc}$. This analysis gives a $T_1$ lifetime of 232 μs for the Spiro-4-DPA system, very similar to the 240 μs estimated for DPA by analyzing the upconverted emission decay of a reference DPA/PtOEP solution (Fig.3C, bottom). The suitability of Spiro-4-DPA as triplets annihilator is further confirmed by the measurement of the solution upconversion quantum efficiency, $\phi_{uc}$, defined as the ratio between the number of upconverted photons emitted and the number of the absorbed one. [11,44] Figure 3D reports the observed $\phi_{uc}$ as a function of the absorbed excitation intensity for the Spiro-4-DPA:PtOEP solution, in direct comparison with the reference DPA:PtOEP (Figs.S10, S11). Due to the bimolecular nature of the intermolecular TTA process, in both samples the $\phi_{uc}$ increases with the excitation intensity until it saturates to a plateau at high powers.[45] With the Spiro-4-DPA as annihilator/emitter the conversion efficiency reaches its maximum value at $\phi_{uc}^{max}$ = 0.21. Notably, this value is only barely lower than the one of the DPA:PtOEP reference solution of 0.25, in agreement with its $\phi_{pl}$.

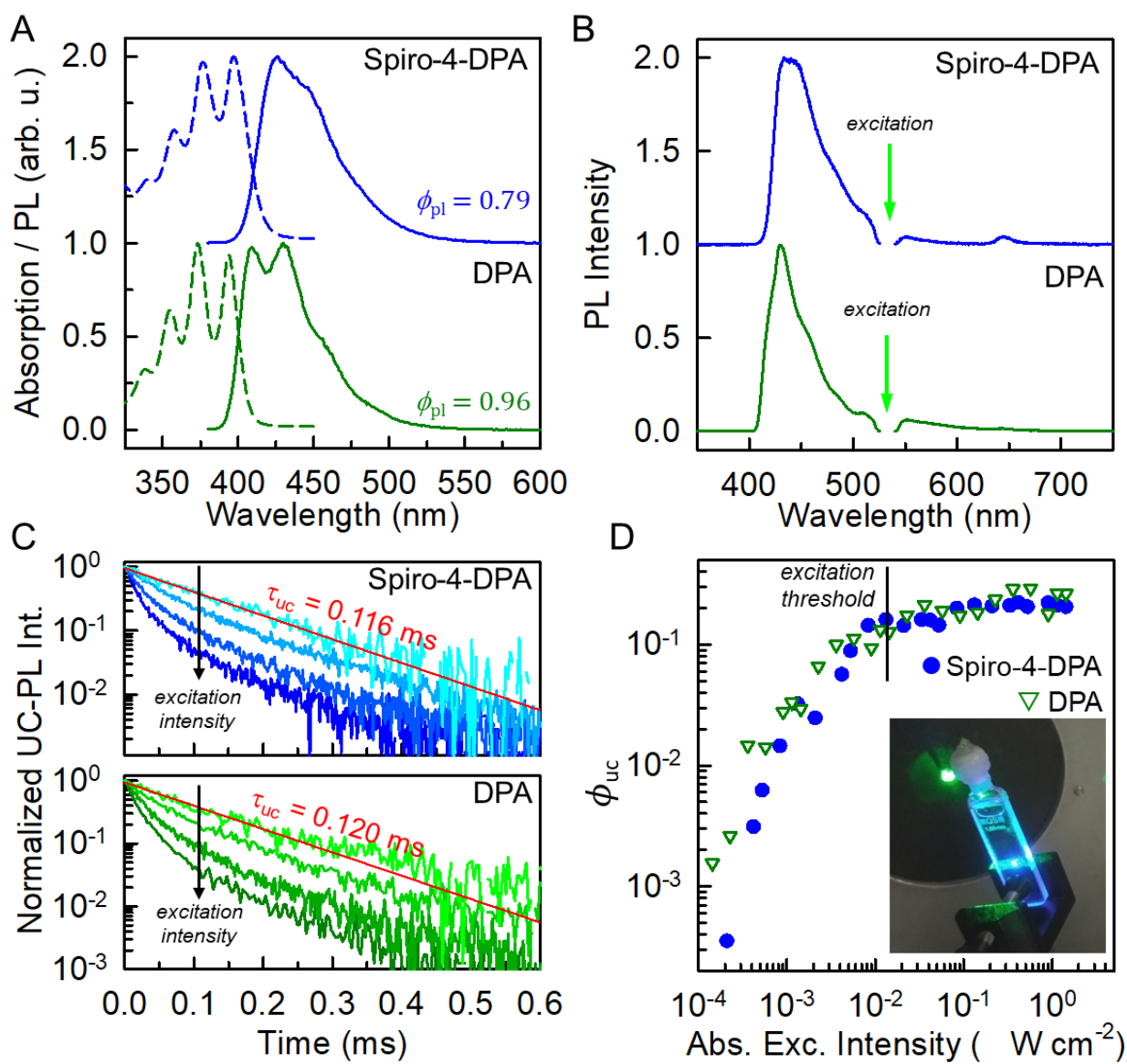

**Fig. 3 | Fluorescence and upconversion properties of Spiro-4-DPA.** (A) Normalized absorption and photoluminescence spectra under *cw* excitation at 350 nm of Spiro-4-DPA and reference DPA solutions ($10^{-6}$ M) in tetrahydrofuran (THF (B) Photoluminescence spectra of Spiro-4-DPA and DPA $10^{-2}$ M solutions in THF with the PtOEP ($10^{-4}$ M) triplet sensitizers under continuous wave 532 nm laser excitation (1 W cm$^{-2}$). The weak residual laser stray light has been removed for clarity. (C) Upconverted photoluminescence (UC-PL) intensity decay with time at 435 under modulate laser excitation at 532 nm as a function of the excitation intensity from $10^{-3}$ W cm$^{-2}$ to 5 W cm$^{-2}$. Solid lines are the fit with single exponential decay function with characteristic decay time $\tau_{uc}$. (D) Upconversion quantum efficiency ($\phi_{uc}$) as a function of the absorbed excitation intensity. The vertical line marks the excitation intensity threshold at which $\phi_{uc}$ is half of its maximum.

The obtained results demonstrate the possibility to use of the Spiro-4-DPA as annihilator/emitter system for *s*TTA-UC. When the annihilator concentration is kept very high ($10^{-4}$ to $10^{-2}$ M) in order to maximize the energy transfer rate and yield, the same dependence of $\phi_{uc}$ vs. the excitation intensity is observed using both Spiro-4-DPA and DPA because in such a conditions the kinetic of the TTA is dominated by the intermolecular process, as usual in bicomponent solutions. (*36*) To directly observe the eventual occurrence of the intramolecular TTA mechanism, we therefore designed a dedicated experiment. The key point is the ability to create simultaneously two triplets on the same emitter molecule. The quantum-mechanical model indicates that this is possible thanks to the Spiro-4-DPA electronic structure, but such result can be evidenced experimentally only in particular conditions. Basically, we must use a concentration of annihilators low enough to make the intermolecular TTA yield negligible. At the same time the concentration of annihilators must be sufficiently high to have some energy transfer, and the excitation intensity must enable the generation of two excitons in the same molecule within their lifetime, thus activating the intramolecular TTA and therefore having chances to detect the resulting upconverted emission. To point out the best sample composition to investigate the intramolecular phenomenon, we calculated the upconversion efficiency of the inter- and intramolecular TTA, taken as two independent processes, as a function of the excitation intensity for a series of solutions with different annihilator concentrations (Supplementary Information). It is worth remembering that the intramolecular TTA obeys to the kinetics of the confined-TTA, [24,46] where the upconversion yield is determined by the binomial distribution of energy among annihilators. On the contrary the intermolecular up conversion is classically controlled by the molecular diffusion with an efficiency which grows linearly with the excitation intensity up to its saturation value. This means that the global upconversion yield dependency on the excitation intensity is completely different in the two cases, thus enabling to experimentally distinguish the two processes. Figure 4A shows the theoretical $\phi_{uc}^*$ values calculated for a series of upconverting solutions with annihilator concentrations ranging from $10^{-9}$ M up to $10^{-6}$ M. The efficiency curves have been plotted as a function of the excitation intensity ranging from $10^{-2}$ to $10^4$ W cm$^{-2}$. The triplet sensitizer concentration is set at $10^{-4}$ M. In agreement with the confined-TTA kinetics, [24] the intramolecular TTA efficiency approaches very quickly the saturation to the maximum value. This is reached approximately at 1 W cm$^{-2}$, independently from the annihilator concentration because its independence on the average distance between the annihilator molecules. Only the maximum yield increases with the Spiro-4-DPA amount following the enhancement of the energy transfer efficiency at higher annihilator amounts (Supplementary Table S3). On the other side, the curves calculated for the intermolecular TTA efficiency in the corresponding samples grow linearly with the excitation intensity, and the power required to reach the maximum efficiency plateau increases by lowering the annihilator concentration which is a direct consequence of the large average distance between two excited molecules. [45] The intra- and intermolecular mechanism behaviors can be therefore distinguished. The calculated curves indeed clearly shows that for annihilator concentrations below $10^{-7}$ M and excitation intensities lower than the intensity threshold the probability to create two triplets on the same molecule is larger than that one of bimolecular collision between two distinct excited annihilators. Thus in this condition the intramolecular process is the dominating annihilation mechanism that could be directly observed. However, at concentration as low as $10^{-9}$ M of Spiro-4-DPA the output upconverted emission signal is expected to be too weak to have reliable measurements. Accordingly, we chose as model system for evidencing the intramolecular TTA a solution with $10^{-8}$ M of Spiro-4-DPA, where the intermolecular TTA is the dominant mechanism in the excitation range between $10^{-2}$ and $10^2$ W cm$^{-2}$ and the energy transfer efficiency is ~ 1% (Supplementary Table S3), and a solution with $10^{-6}$ M of Spiro-4-DPA as model system for the intermolecular TTA.

We investigated such upconverting solutions by steady state photoluminescence spectroscopy. Figure 4B shows the normalized upconversion yield of $10^{-8}$ M vs. $10^{-6}$ M Spiro-4-DPA solutions in THF with $10^{-4}$ PtOEP. In excellent agreement with the calculations (solid line), the efficiency of the $10^{-8}$ M solution increases

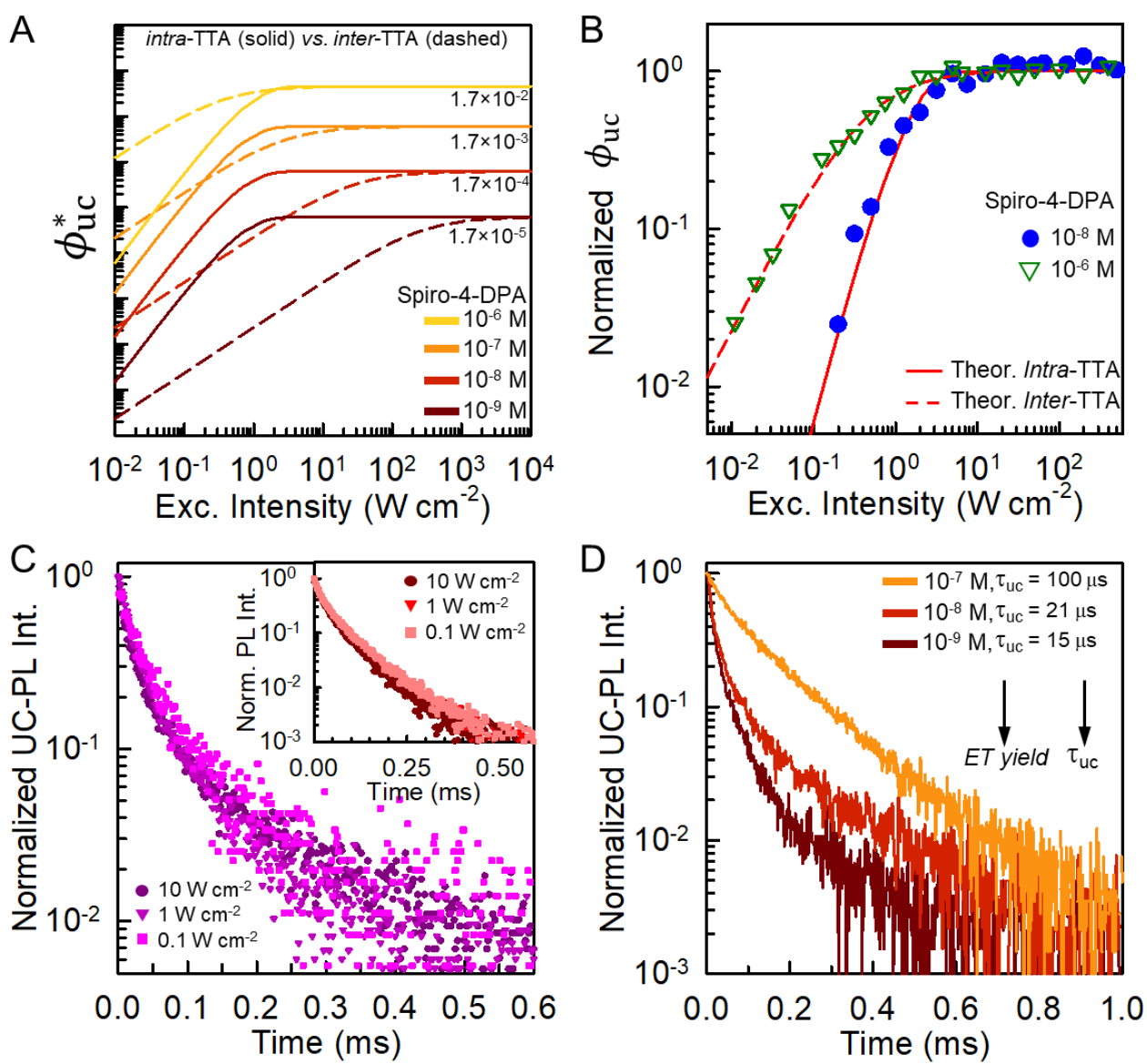

**Fig. 4. Modeling and observation of intramolecular triplet-triplet annihilation (TTA).** (A) Calculated relative efficiencies of intermolecular (*inter-*, dashed line) and intramolecular (*intra-*, solid line) TTA in a series of THF solutions with the triplet sensitizer PtOEP ($10^{-4}$ M, absorptance 0.45 with 0.1 cm optical path) and different concentrations of the annihilator Spiro-4-DPA as a function of the *cw* excitation intensity at 532 nm. (B) Normalized upconverted photoluminescence intensity (UC-PL) under 532 nm *cw* laser excitation for two upconverting THF solutions containing $10^{-8}$ M (dots) and $10^{-6}$ M (triangles) Spiro-4-DPA. The PtOEP concentration is fixed at $10^{-4}$ M. The lines are the fit of data with intra-TTA (solid line) and inter-TTA (dashed line) model functions. (C) UC-PL intensity at 435 nm as a function of time and excitation intensity with $10^{-8}$ M Spiro-4-DPA. Inset: residual phosphorescence intensity from PtOEP at 645 nm as a function of time and excitation for the same sample. (D) UC-PL intensity at 435 nm as a function of time and Spiro-4-DPA concentration under modulated laser excitation at 532 nm (5 W cm$^{-2}$).

accordingly to the intramolecular-TTA dynamics, while the $10^{-6}$ M sample follows the standard behavior of the intermolecular process (dashed line). These finding strongly hints the occurrence of the intramolecular mechanism in the $10^{-8}$ M solution which can be directly demonstrated with time resolved experiments. Figure 4C shows the temporal evolution of the upconverted emission intensity under pulsed laser excitation for the $10^{-8}$ M Spiro-4-DPA solution. In this experiment the excitation intensity ranges from 0.1 W cm$^{-2}$ to 10 W cm$^{-}$

$^2$, thus below the excitation power threshold. In all the measurements the observed up-conversion lifetime is almost constant around 20-30 µs which is a lifetime much shorter than the natural lifetime of the Spiro-4-DPA triplet (Supplementary Table S3). Therefore, despite we are well below the excitation power threshold, the emission lifetime does not change and it does not reflect the long natural lifetime of the triplet states involved in the upconversion process as in the classical intermolecular TTA systems. On the contrary, the observed lifetime perfectly matches the lifetime of the sensitizer PtOEP excited states, the energy reservoir from which the energy transfer takes place to populate the annihilating triplets (inset of Fig.4C). This peculiar behavior can be explained by supposing an intramolecular process in which the upconversion is due to a double excitation of a single Spiro-4-DPA molecules by two sensitizers that transfer their energy within their excited state lifetime. Once the two quanta of energy have been transferred, the annihilation is an instantaneous process (at least in this time range) [47] and it does not require any additional migration step as in the intermolecular TTA. Further support to this picture comes from measurements of the up-conversion lifetime above the excitation intensity threshold at different Spiro-4-DPA concentrations which span from the inter- to the intra-molecular regime (Fig. 4D). The data show that the upconverted emission lifetime, calculated as the time at which the emission intensity is reduced by a factor 1/*e*, shortens by almost one order of magnitude from 100 µs to 15 µs decreasing the emitter amount from $10^{-7}$ M (intermolecular regime) down to $10^{-9}$ M (intramolecular regime). This means that while the rate and yield of the energy transfer from sensitizers is reducing, the TTA rate is increasing, thus accelerating the upconverted emission intensity decay. This behavior can be ascribed only to the occurrence of intramolecular TTA. Indeed, by reducing the energy transfer yield we also increase the mean distance between excited emitters, thus progressively switching off the intermolecular process. This usually implies an increment of the upconverted emission lifetime, because of the reduction of the intermolecular TTA rate. [43] Conversely, in our sample the emission lifetime is shortened, in agreement with the fact that, by eliminating the intermolecular pathway, we can only detect the upconverted photons generated by confined intramolecular TTA, which effective rate is faster that the intermolecular mechanism and usually hindered by the slower processes involved [24]. Considering these latest findings, the full set of time resolved photoluminescence data presented and discussed supports unambiguously the fact that, at the selected low annihilator concentrations, the *s*TTA-UC mechanism is dominated by the intramolecular TTA processes on single Spiro-4-DPA molecules, from which the detected upconverted photons are generated.

**Discussion.**

In summary, we designed and synthetized a new conjugated annihilator/emitter for photon upconversion based on triplet fusion capable of hosting simultaneously more than one triplet exciton, thanks to its finely tuned electronic properties. This allows to achieve and directly observe for the first time a photon upconversion entirely ascribed to an intramolecular TTA process, thus demonstrating the possibility to realize single molecule upconverters for diffusion-free photon managing applications. This observation is a crucial milestone towards the fabrication of new materials for applications in solid state devices. Indeed, the possibility to exploit an intramolecular mechanism can completely change the classical layout of $s$TTA-UC materials, where annihilators are usually taken in large excess with respect to sensitizers, by guiding the development of supramolecular structures including multiple sensitizers (Fig.5) to be coupled with single intramolecular upconverter. This architecture will allow to directly exploit the energy of stored triplets without any need of molecular/excitation diffusion, thus completely surpassing the actual limit of $s$TTA-UC application in the solid state and opening the way to alternative energetic pathways to exploit the energy naturally stored in triplet state of conjugated systems.

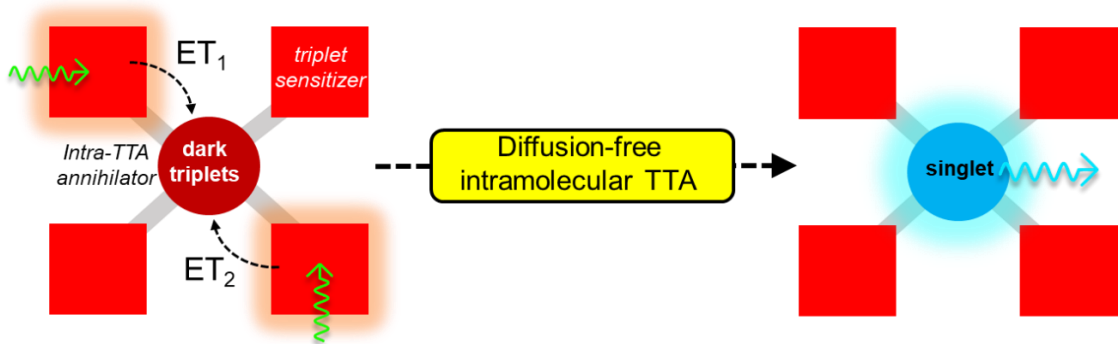

**Fig. 5**. **Photon upconversion in supramolecular systems based on intramolecular TTA mechanism.** Diffusion-free photon upconversion can be achieved by coupling multiple sensitizers to a single annihilator/emitter moiety that shows intramolecular TTA. Sensitizers and emitter are therefore close packed maximizing the energy transfer rate and enabling the creation of upon multiple sequential absorption of green photons.

**Data availability**
The data that support the plots within this paper and other findings of this study are available from the corresponding author upon reasonable request.

# References

1. Einzinger, M. *et al.* Sensitization of silicon by singlet exciton fission in tetracene. *Nature* **571**, 90-94, doi:10.1038/s41586-019-1339-4 (2019).
2. Kim, J. U. *et al.* Nanosecond-time-scale delayed fluorescence molecule for deep-blue OLEDs with small efficiency rolloff. *Nature Communications* **11**, 1765, doi:10.1038/s41467-020-15558-5 (2020).
3. Chan, C.-Y. *et al.* Stable pure-blue hyperfluorescence organic light-emitting diodes with high-efficiency and narrow emission. *Nature Photonics* **15**, 203-207, doi:10.1038/s41566-020-00745-z (2021).
4. Di, D. *et al.* Efficient Triplet Exciton Fusion in Molecularly Doped Polymer Light-Emitting Diodes. *Advanced Materials* **29**, 1605987, doi:https://doi.org/10.1002/adma.201605987 (2017).
5. Wang, X. *et al.* Organic phosphors with bright triplet excitons for efficient X-ray-excited luminescence. *Nature Photonics* **15**, 187-192 (2021).
6. Wang, J.-X. *et al.* Nearly 100% energy transfer at the interface of metal-organic frameworks for X-ray imaging scintillators. *Matter* (2021).
7. Kerzig, C. & Wenger, O. S. Sensitized triplet–triplet annihilation upconversion in water and its application to photochemical transformations. *Chemical Science* **9**, 6670-6678, doi:10.1039/C8SC01829D (2018).
8. Baluschev, S. *et al.* Up-Conversion Fluorescence: Noncoherent Excitation by Sunlight. *Physical Review Letters* **97**, 143903, doi:10.1103/PhysRevLett.97.143903 (2006).
9. Mahboub, M., Huang, Z. & Tang, M. L. Efficient Infrared-to-Visible Upconversion with Subsolar Irradiance. *Nano Letters* **16**, 7169-7175, doi:10.1021/acs.nanolett.6b03503 (2016).
10. Koharagi, M. *et al.* Green-to-UV photon upconversion enabled by new perovskite nanocrystal-transmitter-emitter combination. *Nanoscale*, doi:10.1039/D1NR06588B (2021).
11. Sun, W. *et al.* Highly efficient photon upconversion based on triplet–triplet annihilation from bichromophoric annihilators. *Journal of Materials Chemistry C* **9**, 14201-14208, doi:10.1039/D1TC01569A (2021).
12. Pawlicki, M., Collins, H. A., Denning, R. G. & Anderson, H. L. Two-photon absorption and the design of two-photon dyes. *Angewandte Chemie International Edition* **48**, 3244-3266 (2009).
13. Zhou, J., Liu, Z. & Li, F. Upconversion nanophosphors for small-animal imaging. *Chemical Society Reviews* **41**, 1323-1349 (2012).
14. Kim, H.-i., Kwon, O. S., Kim, S., Choi, W. & Kim, J.-H. Harnessing low energy photons (635 nm) for the production of H2O2 using upconversion nanohybrid photocatalysts. *Energy & Environmental Science* **9**, 1063-1073, doi:10.1039/C5EE03115J (2016).
15. Hill, S. P. & Hanson, K. Harnessing Molecular Photon Upconversion in a Solar Cell at Sub-solar Irradiance: Role of the Redox Mediator. *Journal of the American Chemical Society* **139**, 10988-10991, doi:10.1021/jacs.7b05462 (2017).
16. Pedrini, J. & Monguzzi, A. Recent advances in the application triplet–triplet annihilation-based photon upconversion systems to solar technologies. *Journal of Photonics for Energy* **8**, 022005 (2017).
17. Liu, Q. *et al.* Highly photostable near-IR-excitation upconversion nanocapsules based on triplet–triplet annihilation for in vivo bioimaging application. *ACS applied materials & interfaces* **10**, 9883-9888 (2018).
18. Park, J., Xu, M., Li, F. & Zhou, H.-C. 3D Long-Range Triplet Migration in a Water-Stable Metal–Organic Framework for Upconversion-Based Ultralow-Power in Vivo Imaging. *Journal of the American Chemical Society* **140**, 5493-5499, doi:10.1021/jacs.8b01613 (2018).
19. Sasaki, Y. *et al.* Near-Infrared Optogenetic Genome Engineering Based on Photon-Upconversion Hydrogels. *Angewandte Chemie* **131**, 17991-17997 (2019).
20. Hagstrom, A. L. *et al.* Flexible and Micropatternable Triplet–Triplet Annihilation Upconversion Thin Films for Photonic Device Integration and Anticounterfeiting Applications. *ACS applied materials & interfaces* **10**, 8985-8992 (2018).
21. Huang, L., Le, T., Huang, K. & Han, G. Enzymatic enhancing of triplet–triplet annihilation upconversion by breaking oxygen quenching for background-free biological sensing. *Nature Communications* **12**, 1898, doi:10.1038/s41467-021-22282-1 (2021).
22. Wang, W. *et al.* Efficient Triplet–Triplet Annihilation-Based Upconversion for Nanoparticle Phototargeting. *Nano Letters* **15**, 6332-6338, doi:10.1021/acs.nanolett.5b01325 (2015).

23   Perego, J. *et al.* Engineering Porous Emitting Framework Nanoparticles with Integrated Sensitizers for Low-Power Photon Upconversion by Triplet Fusion. *Advanced Materials* **31**, 1903309, doi:https://doi.org/10.1002/adma.201903309 (2019).

24   Meinardi, F. *et al.* Quasi-thresholdless Photon Upconversion in Metal–Organic Framework Nanocrystals. *Nano Letters* **19**, 2169-2177, doi:10.1021/acs.nanolett.9b00543 (2019).

25   Roy, I. *et al.* Photon Upconversion in a Glowing Metal–Organic Framework. *Journal of the American Chemical Society* **143**, 5053-5059, doi:10.1021/jacs.1c00298 (2021).

26   Simon, Y. C. & Weder, C. Low-power photon upconversion through triplet–triplet annihilation in polymers. *Journal of Materials Chemistry* **22**, 20817-20830, doi:10.1039/C2JM33654E (2012).

27   Peng, J., Guo, X., Jiang, X., Zhao, D. & Ma, Y. Developing efficient heavy-atom-free photosensitizers applicable to TTA upconversion in polymer films. *Chemical Science* **7**, 1233-1237, doi:10.1039/C5SC03245H (2016).

28   Kamada, K. *et al.* Efficient triplet–triplet annihilation upconversion in binary crystalline solids fabricated via solution casting and operated in air. *Materials Horizons* **4**, 83-87 (2017).

29   Hill, S. P., Dilbeck, T., Baduell, E. & Hanson, K. Integrated Photon Upconversion Solar Cell via Molecular Self-Assembled Bilayers. *ACS Energy Letters* **1**, 3-8, doi:10.1021/acsenergylett.6b00001 (2016).

30   Oldenburg, M. *et al.* Photon Upconversion at Crystalline Organic–Organic Heterojunctions. *Advanced Materials* **28**, 8477-8482, doi:https://doi.org/10.1002/adma.201601718 (2016).

31   Izawa, S. & Hiramoto, M. Efficient solid-state photon upconversion enabled by triplet formation at an organic semiconductor interface. *Nature Photonics* **15**, 895-900, doi:10.1038/s41566-021-00904-w (2021).

32   Saenz, F. *et al.* Nanostructured Polymers Enable Stable and Efficient Low-Power Photon Upconversion. *Advanced Functional Materials* **31**, 2004495, doi:https://doi.org/10.1002/adfm.202004495 (2021).

33   Dzebo, D., Börjesson, K., Gray, V., Moth-Poulsen, K. & Albinsson, B. Intramolecular Triplet–Triplet Annihilation Upconversion in 9,10-Diphenylanthracene Oligomers and Dendrimers. *The Journal of Physical Chemistry C* **120**, 23397-23406, doi:10.1021/acs.jpcc.6b07920 (2016).

34   Edhborg, F., Bildirir, H., Bharmoria, P., Moth-Poulsen, K. & Albinsson, B. Intramolecular Triplet–Triplet Annihilation Photon Upconversion in Diffusionally Restricted Anthracene Polymer. *The Journal of Physical Chemistry B* **125**, 6255-6263, doi:10.1021/acs.jpcb.1c02856 (2021).

35   Olesund, A., Gray, V., Mårtensson, J. & Albinsson, B. Diphenylanthracene Dimers for Triplet–Triplet Annihilation Photon Upconversion: Mechanistic Insights for Intramolecular Pathways and the Importance of Molecular Geometry. *Journal of the American Chemical Society* **143**, 5745-5754, doi:10.1021/jacs.1c00331 (2021).

36   Cravcenco, A. *et al.* Multiplicity conversion based on intramolecular triplet-to-singlet energy transfer. *Science Advances* **5**, eaaw5978, doi:doi:10.1126/sciadv.aaw5978 (2019).

37   Dreuw, A. & Head-Gordon, M. Single-reference ab initio methods for the calculation of excited states of large molecules. *Chemical reviews* **105**, 4009-4037 (2005).

38   Furche, F. *et al.* Circular dichroism of helicenes investigated by time-dependent density functional theory. *Journal of the American Chemical Society* **122**, 1717-1724 (2000).

39   Neese, F. WIREs Comput. Mol. Sci. 2012, 2, 73–78; b) F. Neese. *WIREs Comput. Mol. Sci* **8**, e1327 (2018).

40   Neese, F. Software update: the ORCA program system, version 4.0. *Wiley Interdisciplinary Reviews: Computational Molecular Science* **8**, e1327 (2018).

41   Cotton, F. A. *Chemical applications of group theory*.  (John Wiley & Sons, 1991).

42   Inokuti, M. & Hirayama, F. Influence of Energy Transfer by the Exchange Mechanism on Donor Luminescence. *The Journal of Chemical Physics* **43**, 1978-1989, doi:10.1063/1.1697063 (1965).

43   Pope, M. S. C. E. *Electronic processes in organic crystals and polymers*.  (Oxford University Press, 1999).

44   Zhou, Y., Castellano, F. N., Schmidt, T. W. & Hanson, K. On the Quantum Yield of Photon Upconversion via Triplet–Triplet Annihilation. *ACS Energy Letters* **5**, 2322-2326, doi:10.1021/acsenergylett.0c01150 (2020).

45   Monguzzi, A., Mezyk, J., Scotognella, F., Tubino, R. & Meinardi, F. Upconversion-induced fluorescence in multicomponent systems: Steady-state excitation power threshold. *Physical Review B* **78**, 195112, doi:10.1103/PhysRevB.78.195112 (2008).

46  Ronchi, A. & Monguzzi, A. Developing solid-state photon upconverters based on sensitized triplet–triplet annihilation. *Journal of Applied Physics* **129**, 050901, doi:10.1063/5.0034943 (2021).
47  Ronchi, A. *et al.* High Photon Upconversion Efficiency with Hybrid Triplet Sensitizers by Ultrafast Hole-Routing in Electronic-Doped Nanocrystals. *Advanced Materials* **32**, 2002953 (2020).

**Acknowledgments.**
SM, GM, FM, LB and AM gratefully acknowledge the Italian Ministry of University (MUR) for financial support through Grant Dipartimenti di Eccellenza-2017 "Materials for Energy" and Grant PRIN-2017 BOOSTER (2017YXX8AZ) is gratefully acknowledged.

**Author contributions.**
Materials design: GM, LB. Materials synthesis and haracterization: SMa, SMe, AC, FP. Computational molecular modelling: GM. Numerical kinetic modeling: AR, FM, AM. Spectroscopy experiments: AR, AM. Writing: GM, FM, LB, AM. Project design, and supervision: FM, LB, AM.

**Competing interests.**
Authors declare that they have no competing interests.

**Additional information**
Supplementary information including Materials and Methods details and additional data are available in the online version of the paper. Correspondence and requests for materials should be addressed to A.M.